\documentclass[11pt,superscriptaddress,aps]{revtex4}
\usepackage{graphicx}

\begin{document}
\title{Implication of the proton electric FF space-like behavior puzzle in various
physical phenomena}
\date{\today}

\author{\underline{A.Z.Dubni\v ckov\'a}}
\address{Dept. of Theoretical Physics, Comenius Univ., Bratislava,
Slovak Republic}

\author{S.Dubni\v cka}
\address{Institute of Physics, Slovak Academy of Sciences,
Bratislava, Slovak Republic}

\vspace{2cm}

\begin{abstract}
By means of the 10-resonance unitary and analytic model of nucleon
electromagnetic structure it is demonstrated that the JLab proton
polarization data on the ratio $\mu_p G_{Ep}(Q^2)/G_{Mp}(Q^2)$ are
consistent with all form factor properties, however, they strongly
require an existence of the zero in $G_{Ep}(Q^2)$ around $Q^2=13
GeV^2$. As a result there are two contradicting behaviors of
$G_{Ep}(Q^2)$ in space-like region. Consequences of this phenomenon
on the charge distribution within the proton, on the saturation of
the new proton-neutron $q^2$-dependent sum rule, on the behavior of
strange nucleon form factors and the deuteron elastic structure
functions through the impulse approximation are investigated.
\end{abstract}

\maketitle

\section{Introduction}

In the framework of the naive quark model the proton is compound of
three-quarks and in electromagnetic (EM) interactions they manifest
the proton (equally well the neutron) EM structure.

   As a result, one doesn't know explicit form of the nucleon matrix
element of the EM current
\begin{eqnarray}
J^{EM}_\mu=2/3\bar u\gamma_\mu u-1/3\bar d\gamma_\mu d-1/3\bar
s\gamma_\mu s.
\end{eqnarray}
Then EM form factors (FF), two independent scalar functions of one
variable $t=-Q^2$ (the squared four-momentum transferred by the
exchanged virtual photon) are introduced to represent the proton EM
structure.

There is some freedom in the choice of them.

The most natural is an introduction of Dirac $F_{1p}(t)$ and Pauli
$F_{2p}(t)$ FF's
\begin{eqnarray}
<p\mid J^{EM}_\mu \mid p>=\bar u(p')\{\gamma_\mu F_{1p}(t)+i
\frac{\sigma_{\mu\nu}q_\nu}{2m^2_p}F_{2p}(t)\}u(p).
\end{eqnarray}

   The most suitable in the extraction of experimental information on the proton EM structure are
Sachs electric $G_{Ep}(t)$ and magnetic $G_{Mp}(t)$ FFs of the
proton
\begin{eqnarray}
 G_{Ep}(t)=F_{1p}(t)+\frac{t}{4m^2_p}F_{2p}(t)\\\nonumber
 G_{Mp}(t)=F_{1p}(t)+F_{2p}(t)
\end{eqnarray}
giving in the Breit frame the charge and magnetization distributions
within the proton, respectively.

   However, for a construction of models of proton EM
structure the iso-scalar and iso-vector Dirac and Pauli FF's are the
most appropriate, which are defined by the relations
\begin{eqnarray}
<N\mid J^{I=0}_\mu \mid N>=\bar u(p')\{\gamma_\mu F^{I=0}_{1}(t)+i
\frac{\sigma_{\mu\nu}q_\nu}{2m^2_p}F^{I=0}_{2}(t)\}u(p)\\
J^{I=0}_\mu=\frac{1}{6}(\bar u\gamma_\mu u+\bar d\gamma_\mu
d)-\frac{1}{3}\bar s\gamma_\mu s
\end{eqnarray}
and
\begin{eqnarray}
<N\mid J^{I=1}_\mu \mid N>=\bar u(p')\{\gamma_\mu F^{I=1}_{1}(t)+i
\frac{\sigma_{\mu\nu}q_\nu}{2m^2_p}F^{I=1}_{2}(t)\}u(p)\\
J^{I=1}_\mu=\frac{1}{2}(\bar u\gamma_\mu u-\bar d\gamma_\mu d).
\end{eqnarray}

   Iso-scalar and iso-vector Dirac and Pauli FFs, as one can see
from the expressions
\begin{eqnarray}
\nonumber G_{Ep}(t)&=&
[F_1^{I=0}(t)+ F_1^{I=1}(t)]+\frac{t}{4m_p^2}[F_2^{I=0}(t)+F_2^{I=1}(t)];\\
 G_{Mp}(t)&=&[F_1^{I=0}(t)+F_1^{I=1}(t)]+
[F_2^{I=0}(t)+F_2^{I=1}(t)];\\
\nonumber G_{En}(t)&=&
[F_1^{I=0}(t)-F_1^{I=1}(t)]+\frac{t}{4m_n^2}[F_2^{I=0}(t)-F_2^{I=1}(t)];\\
\nonumber G_{Mn}(t)&=&[F_1^{I=0}(t)-F_1^{I=1}(t)]+
[F_2^{I=0}(t)-F_2^{I=1}(t)],
\end{eqnarray}
are  related separately neither to proton, nor to neutron, but to
the nucleons generally. So, one has always to analyze both, proton
and neutron, existing experimental data sets by constructed models
simultaneously.

\section{ Experimental information on nucleon EM form factors}

Between the discovery of proton EM structure in the middle of the
1950's till 2000, abundant proton EM FF data (from DESY, SLAC and
Bonn) in the space-like region ($t<0$) appeared (see Fig.1).

\begin{figure}[htb]
\centerline{\includegraphics[width=0.45\textwidth]{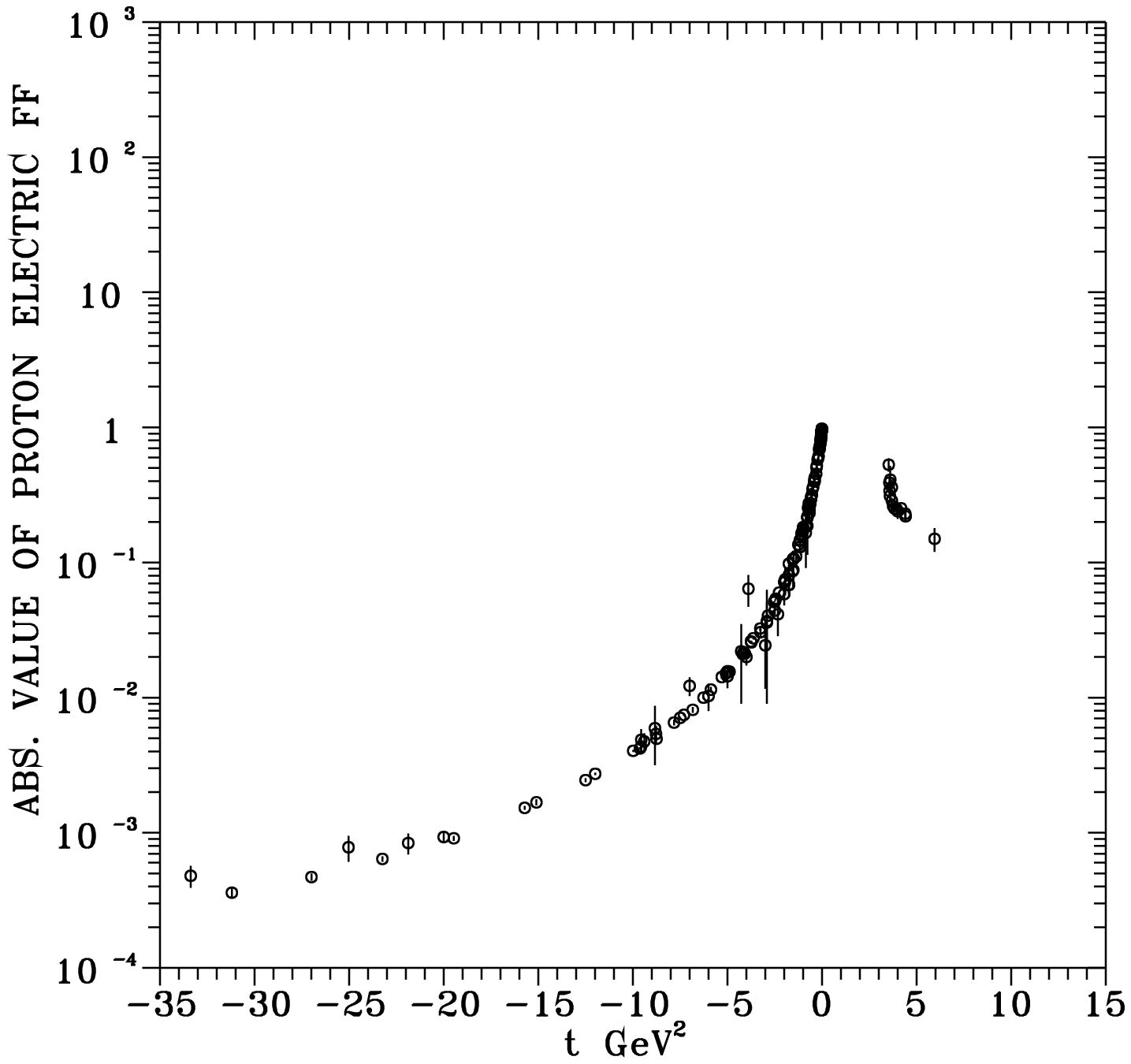}
\qquad
\includegraphics[width=0.45\textwidth]{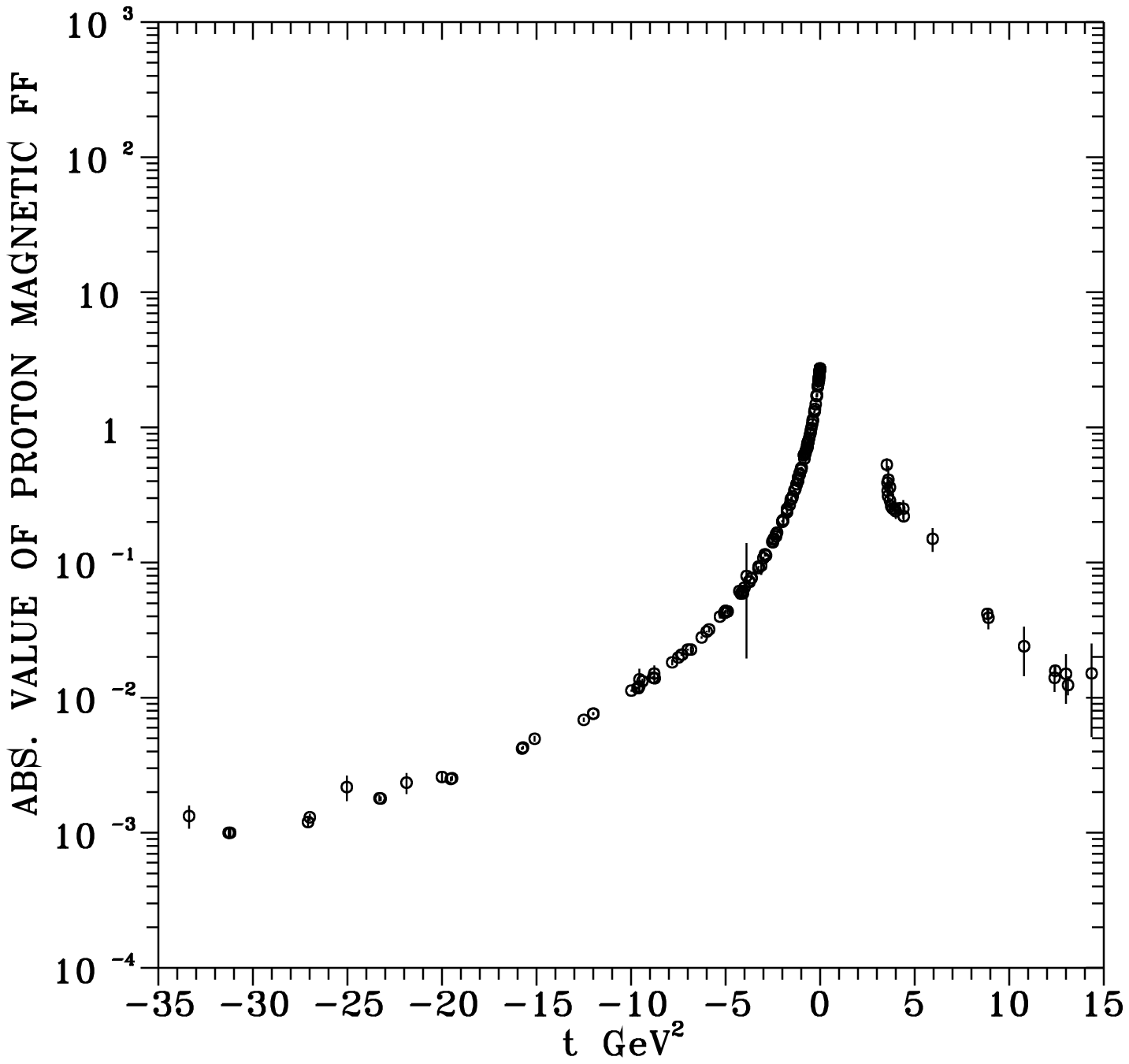}}
\caption{Experimental data on proton electric and magnetic form
factors.}
\end{figure}

They have been obtained from the measured cross section of the
elastic scattering of unpolarized electrons on unpolarized protons
in the laboratory reference frame
\begin{eqnarray}
\nonumber \frac{d\sigma^{lab}(e^-p\to
e^-p)}{d\Omega}=\frac{\alpha^2}{4E^2}\frac{\cos^2(\theta/2)}{\sin^4(\theta/2)}
\frac{1}{1+(\frac{2E}{m_p})\sin^2(\theta/2)}.\nonumber
\end{eqnarray}
\begin{equation}
\left[A(t)+B(t)\tan^2(\theta/2)\right],
\end{equation}
where $\alpha=1/137$, $E$-the incident electron energy
\begin{eqnarray}
 A(t)=\frac{G^2_{Ep}(t)-\frac{t}{4m_p^2}G^2_{Mp}(t)}{1-\frac{t}{4m_p^2}},\\
 B(t)=-2\frac{t}{4m_p^2}G^2_{Mp}(t)
\end{eqnarray}
by  Rosenbluth technique.

One can see from the previous formulas, that the proton magnetic FF
is multiplied by $-t/(4m^2_p)$ factor, i.e. as $-t$ increases, the
measured cross-section (9) becomes dominant by $G^2_{Mp}(t)$ part
contribution, making the extraction of $G^2_{Ep}(t)$ more and more
difficult. So, one can have a confidence only in the proton magnetic
FF data obtained by the Rosenbluth technique.

   By a slightly more complicated method the neutron electric and magnetic FF's
data have been obtained as presented in Fig.2

\begin{figure}[htb]
\centerline{\includegraphics[width=0.45\textwidth]{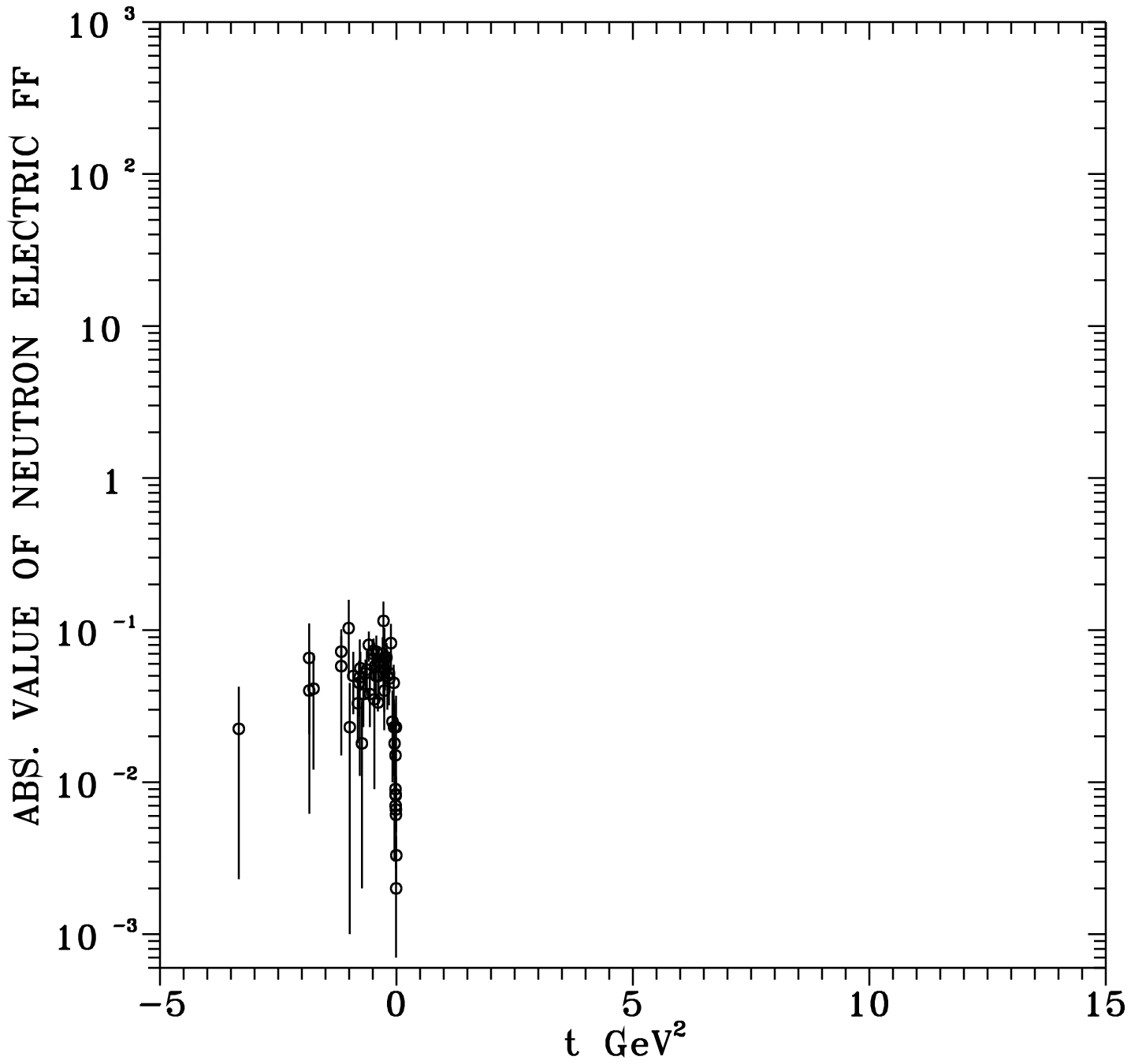}
\qquad
\includegraphics[width=0.45\textwidth]{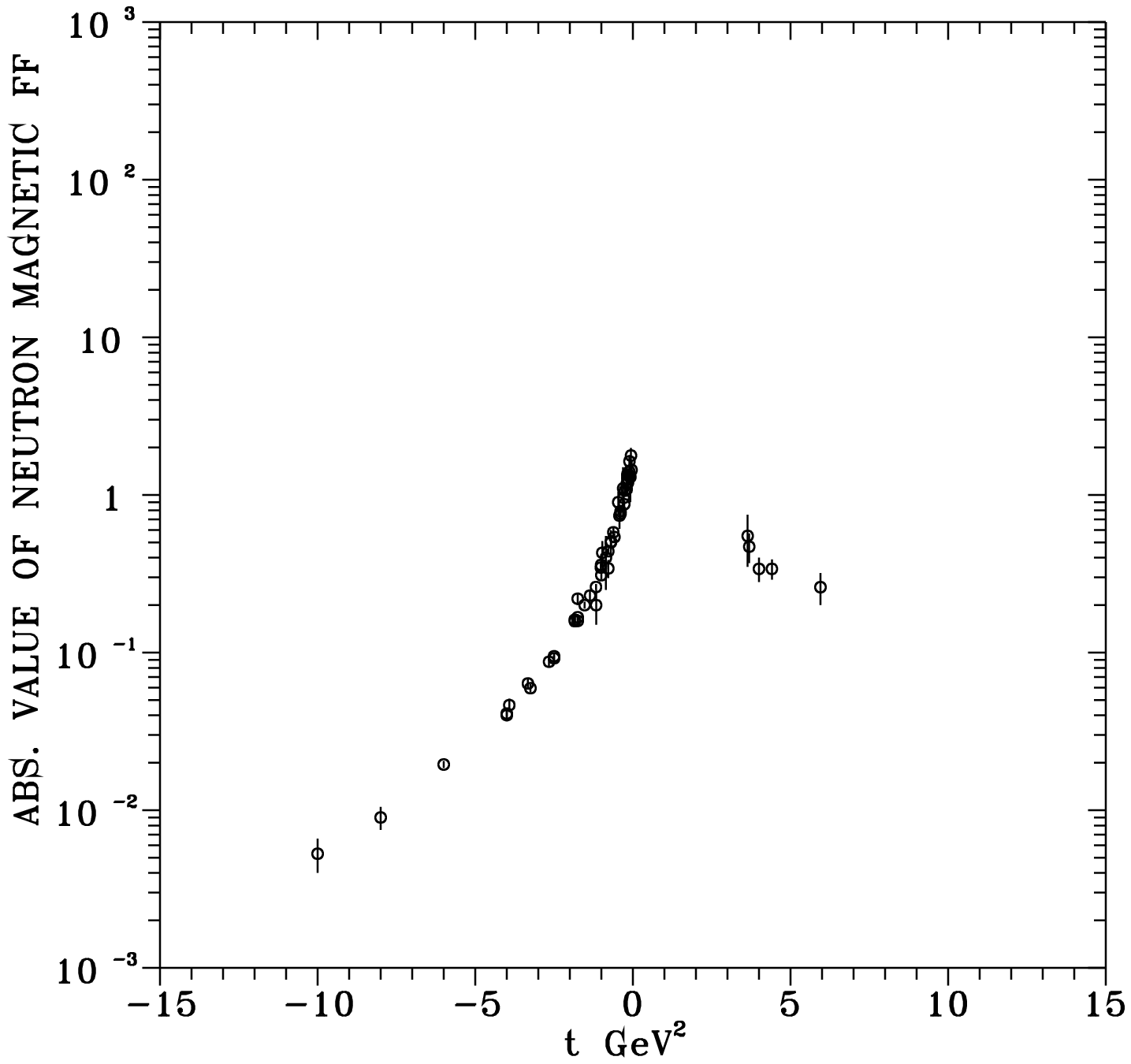}}
\caption{Experimental data on neutron electric and magnetic form
factors.}
\end{figure}

   References on the experimental data on proton and neutron
electromagnetic FF's, obtained by Rosenbluth technique, can be found
in \cite{Dub1}.

More recently at Jefferson Lab \cite{Jon}, \cite{Gay} measuring
simultaneously transverse
\begin{eqnarray}
P_t=\frac{h}{I_0}(-2)\sqrt{\tau(1+\tau)}G_{Mp}G_{Ep}\tan(\theta/2)
\end{eqnarray}
and longitudinal
\begin{eqnarray}
P_l=\frac{h(E+E')}{I_0m_p}\sqrt{\tau(1+\tau)}G^2_{Mp}\tan^2(\theta/2)
\end{eqnarray}
components of the recoil proton's polarization in the electron
scattering plane of the polarization transfer process
$\overrightarrow{e}^{-}p\rightarrow e^{-}\overrightarrow{p}$,
where $h$ is the electron beam helicity, $I_{0}$ is the
unpolarized cross-section excluding $\sigma _{Mott}$ and $\tau
=Q^{2}/4m_{p}^{2}$, the data on the ratio
\begin{eqnarray}
 G_{Ep}/G_{Mp}=-\frac{P_t}{P_l}\frac{(E+E')}{2m_p}\tan(\theta/2)
\end{eqnarray}
were obtained.

\begin{figure}[htb]
\centerline{\includegraphics[width=0.55\textwidth]{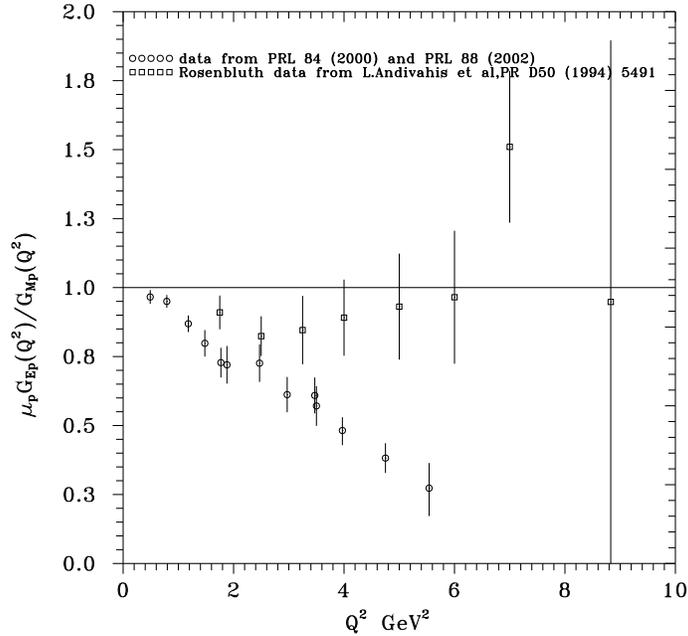}}\caption{
JLab polarization data on the ratio $\mu_{p}G_{Ep}(t)/G_{Mp}(t)$}
\end{figure}

They are in strong disagreement with data obtained by Rosenbluth
technique.

   Taking into account the dominance of $G^p_M(t)$ in the
unpolarized cross-section, we already conjecture the behavior of
$G^p_E(t)$ is responsible for the appeared discrepancy.

This conclusion is supported also in the analysis by our Unitary \&
Analytic (U\&A) model of nucleon electromagnetic structure
\cite{Dub1}.

\section{ Results of analyzes by 10-resonance U\&A model of nucleon
EM structure}

We have achieved simultaneous description of all existing proton and
neutron FF data in space-like and time-like regions by 10-resonance
U\&A model of nucleon EM structure \cite{Dub1} formulated in the
language of iso-scalar $F^{s}_{1,2}(t)$ and iso-vector
$F^{v}_{1,2}(t)$ Dirac and Pauli FFs, saturating them by $\omega,
\phi, \omega', \omega'', \phi'$ and $\rho, \rho', \rho'', \rho''',
\rho''''$, respectively.

The model comprises all known nucleon FF properties
\begin{itemize}
\item
experimental fact of a creation of unstable vector meson resonances
in electron-positron annihilation processes into hadrons
\item
analytic properties of FFs
\item
reality conditions
\item
unitarity conditions
\item
normalizations
\item
asymptotic behaviors as predicted by the quark model of hadrons.
\end{itemize}

First, the analysis of all proton and neutron data obtained by
Rosenbluth technique, together with all proton and neutron data in
time-like region were carried out.

Then, all $G_{Ep}(t)$ space-like data obtained by Rosenbluth
technique were substituted for the new JLab proton polarization data
on the ratio $\mu _{p}G_{Ep}(Q^{2})/G_{Mp}(Q^{2})$ in the interval
 $0.49GeV^{2}\leq Q^{2}\leq 5.54GeV^{2}$ and analyzed together
with all electric proton time-like data and all space-like and
time-like magnetic proton, as well as electric and magnetic neutron,
data \cite{Dub2},\cite{Ada}.

The results are presented in Figs.4 and 5.

\begin{figure}[htb]
\centerline{\includegraphics[width=0.45\textwidth]{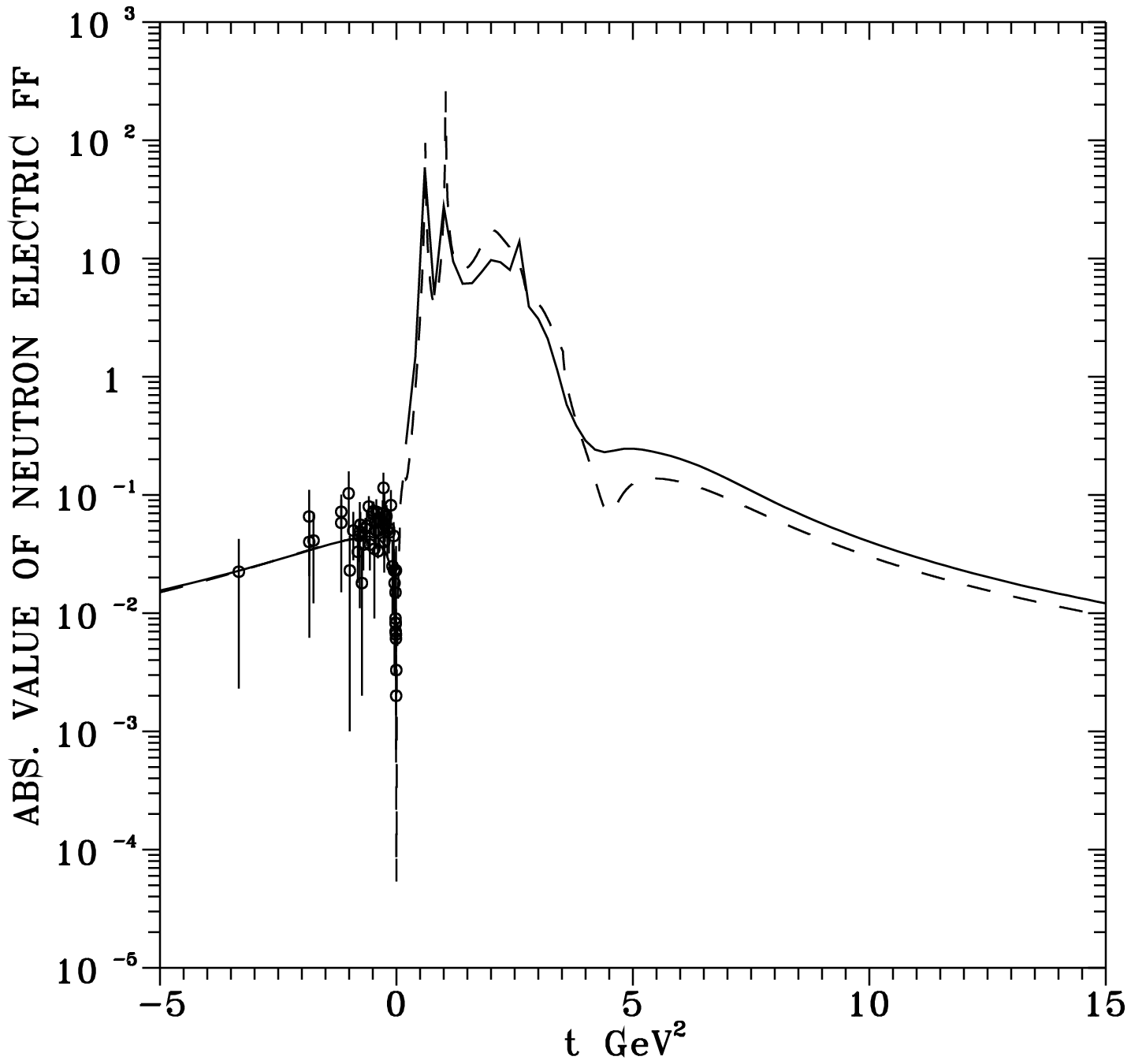}
\qquad
\includegraphics[width=0.45\textwidth]{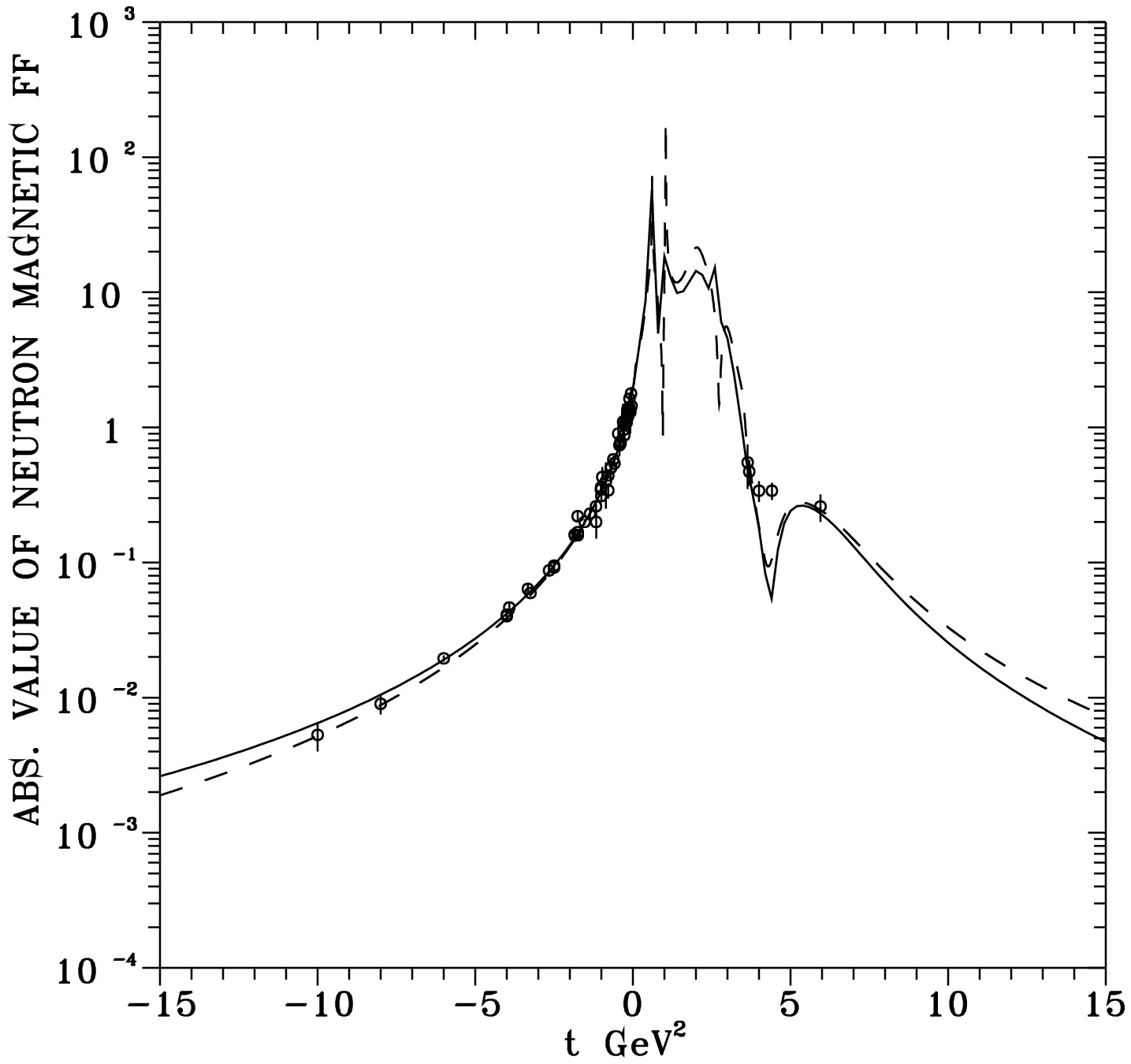}}
\caption{Theoretical behavior of neutron electric and magnetic
form factors.}
\end{figure}
\begin{figure}[htb]
\centerline{\includegraphics[width=0.45\textwidth]{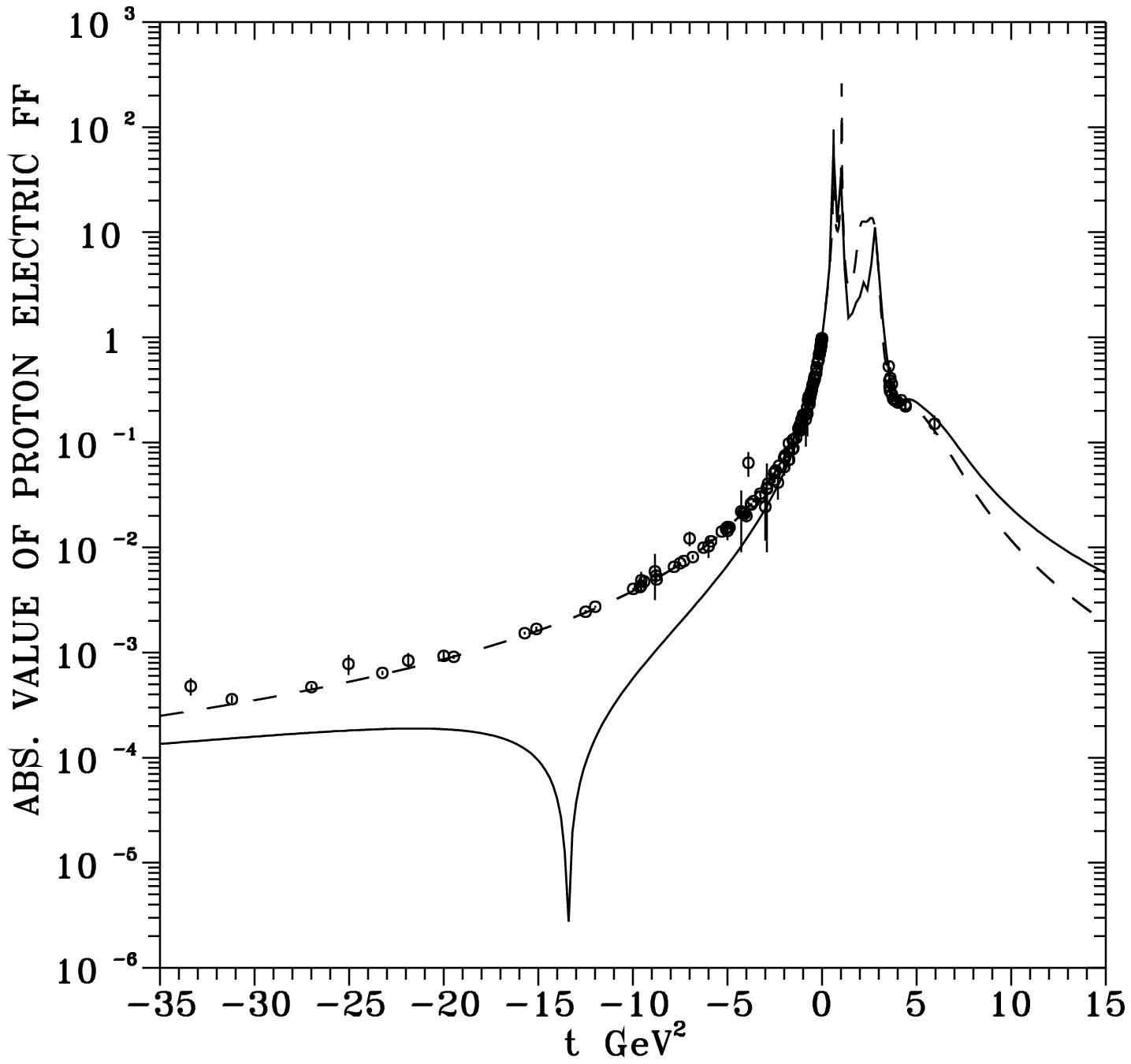}
\qquad
\includegraphics[width=0.45\textwidth]{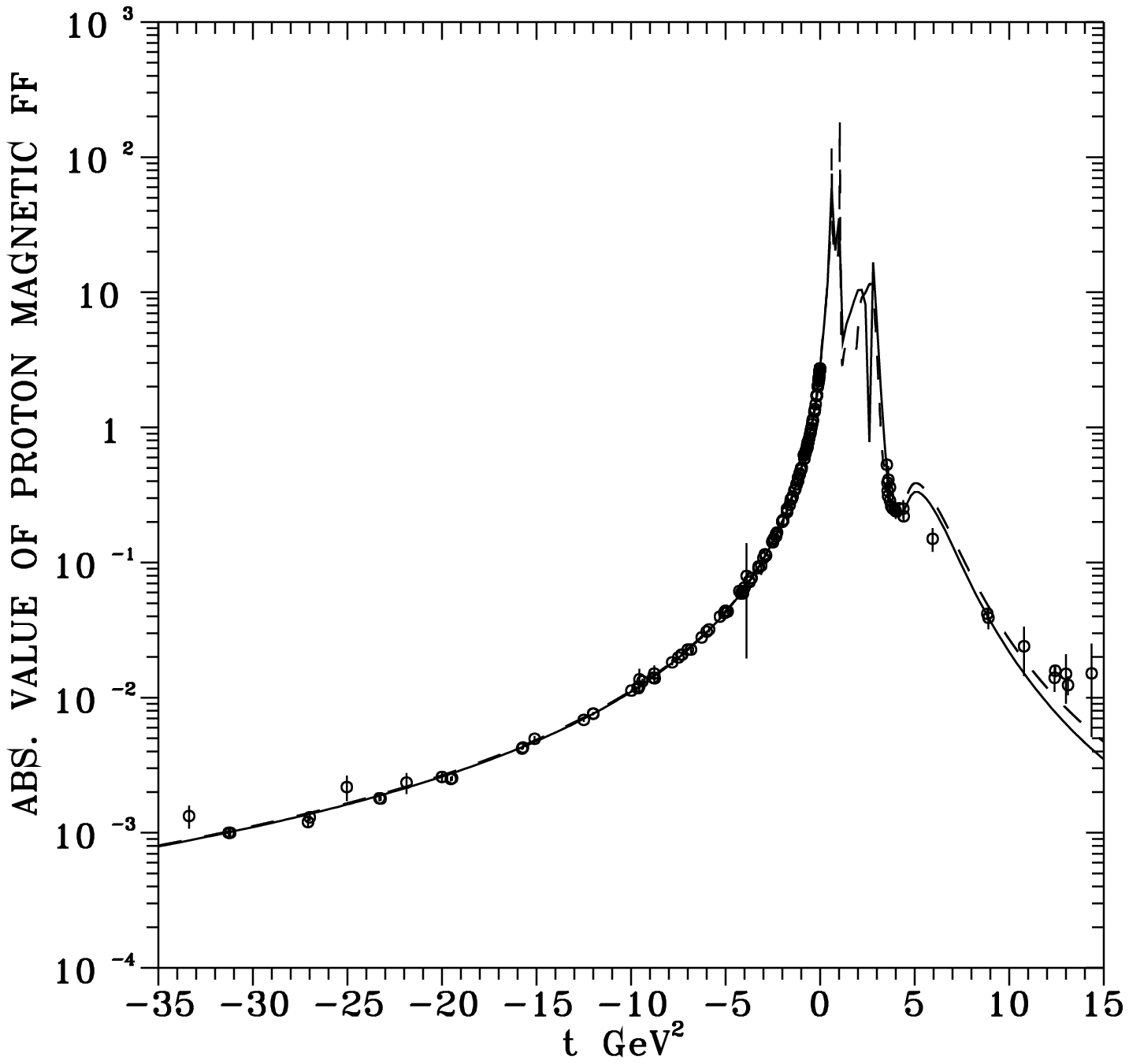}}
\caption{Theoretical behavior of proton electric and magnetic form
factors.}
\end{figure}

From these Figures two consequences follow:
\begin{itemize}
\item
   The fact, that almost nothing is changed in a description of
 $G_{En}(t)$, $G_{Mn}(t)$ and $G_{Mp}(t)$ in both, the space-like
and time-like regions, and also $|G_{Ep}(t)|$ in the time-like
region, supports our hypothesis, that the discrepancy between the
calculated old and measured new ratios $G_{Ep}(t)/G_{Mp}(t)$ is
really created by different behaviors of $G_{Ep}(t)$.
\item
   The new behavior of $G_{Ep}(t)$ (the full line in Fig.5) extracted
from the JLab polarization data on $G_{Ep}(t)/G_{Mp}(t)$ is
consistent with all known FF properties, including also the
asymptotic behavior, but strongly requires an existence of FF zero
around $t=-13 GeV^2$.
\end{itemize}

As a result of our analysis there are two sets of nucleon FF data
differing by $G_{Ep}(t)$ behavior in $t<0$ region.

We would like to note, that the expressions for
$\frac{d\sigma^{lab}(e^-p\to e^-p)}{d\Omega}$ and $P_t$, $P_l$ are
calculated in the one photon exchange approximation to be justified
theoretically.

\section{Attempts to solve the problem}

Despite the fact, that the one photon exchange approximation is
justified theoretically, there were attempts to solve the problem by
inclusion of additional radiative correction terms, related to
two-photon exchange approximations
\cite{Gui},\cite{Blu},\cite{Chen},\cite{Rek},\cite{Dub3}.

The analysis revealed:
\begin{itemize}
\item
 the two-photon exchange has a much smaller effect on the
polarization transfer than on the Rosenbluth extractions
\item
the size of the two-photon exchange correction is less than
half the size necessary to explain discrepancy
\end{itemize}

then the problem is still open, though JLab proton
polarization data seem to be more reliable.

\section{Consequences on charge distribution within proton}

The proton charge distribution (assuming to be spherically
symmetric) is an inverse Fourier transform of the proton
electric FF
\begin{equation}
\rho _{p}(r)=\frac{1}{(2\pi )^{3}}\int e^{-iQr}G_{Ep}(Q^{2})d^{3}Q
\end{equation}
from where
\begin{equation}
\rho _{p}(r)=\frac{4\pi }{(2\pi )^{3}}\int_{0}^{\infty }G_{Ep}(Q^{2})\frac{%
\sin (Qr)}{Qr}Q^{2}dQ.
\end{equation}

Substituting for the $G_{Ep}(Q^{2})$ under the integral:
\begin{itemize}
\item
either the Rosenbluth behavior
\item
or the JLab polarization behavior of  $G_{Ep}(Q^{2})$ with the zero
\end{itemize}
one gets different charge distributions within the proton
given in Fig.6 by dashed and full lines, respectively.

\begin{figure}[htb]
\centerline{\includegraphics[width=0.55\textwidth]{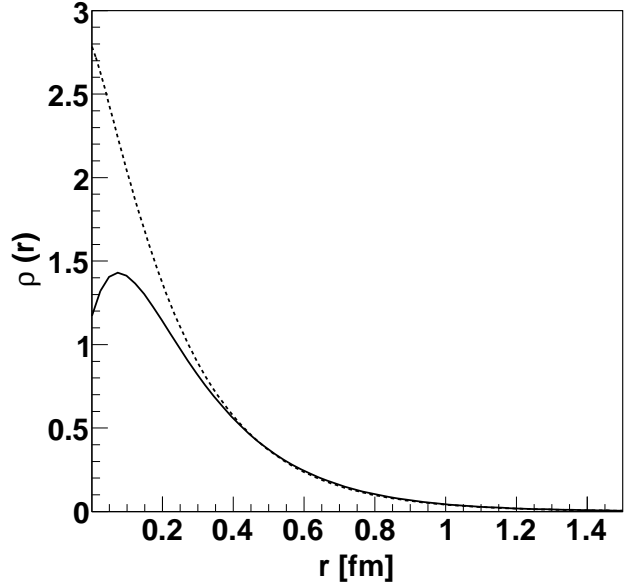}}
\caption{Charge distribution behavior within the proton}
\end{figure}

That all leads also to different mean square proton charge
radii. The old proton charge radius takes the value
$\left\langle r_{p}^{2}\right\rangle =0.68$fm$^{2}$. If JLab
proton polarization data are correct then the new
charge radius $\left\langle r_{p}^{2}\right\rangle =0.72$fm$^{2}$
is larger.

\section{Possible insight by DIS}

   So, the Jlab proton polarization data strongly require an existence
of the zero, i.e. the diffraction minimum in the space-like
region of $|G_{Ep}(t)|$ around $t=-Q^2= 13 GeV^2$.

  Is really the new predicted $t<0$ behavior of $G_{Ep}(t)$ in $t < 0$ correct ?

  It seems to us that this  question could be verified also by DIS, using
 the new sum rule \cite{Bar}
\begin{eqnarray}\label{eq:23}
F_{1p}^2(-{\bf q}^2)+\frac{{\bf q}^2}{4m_p^2} F_{2p}^2(-{\bf q}^2)
-F_{1n}^2(-{\bf q}^2)-\frac{{\bf q}^2}{4m_n^2} F_{2n}^2(-{\bf q}^2)=
 1 - 2\frac{({\bf q}^2)^2}{\pi\alpha^2} \left(\frac{d\sigma^{e^-p
\to e^-X}}{d{\bf q}^2}-\frac{d\sigma^{e^-n \to e^-X}}{d{\bf
q}^2}\right),
\end{eqnarray}
giving into a  relation:
\begin{itemize}
\item
nucleon electromagnetic form factors
\item
with difference of deep inelastic electron-proton and
electron-neutron differential cross-sections.
\end{itemize}

 By measurements of the right hand side of (17) the  true
$t<0$ behavior of the electric proton FF could be chosen

\begin{figure}[htp] 
\centering
\includegraphics[scale=.55]{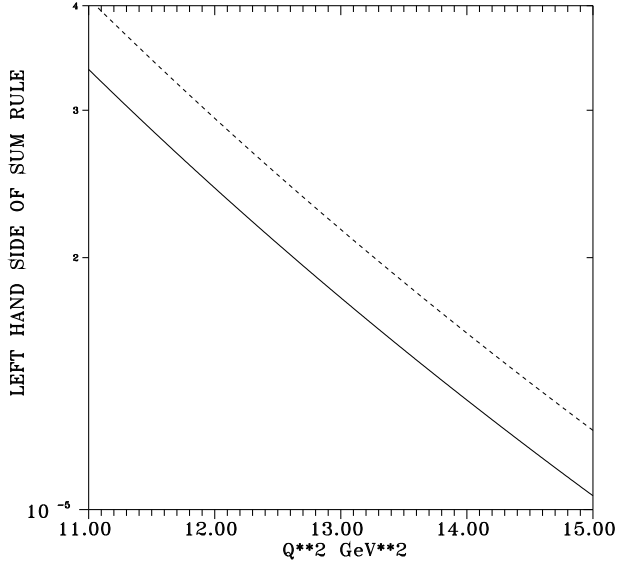}
\caption{$Q^2$ square dependence sum rules results}\label{fig:8}
\end{figure}

\section{ Sensitivity of strange nucleon FFs from two different
behaviors  of $G^p_E(Q^2)$}

The momentum dependence of the nucleon matrix element of the
strange-quark vector current $J^{s}_{\mu}$=$\bar{s}\gamma_{\mu} s$
is contained in the Dirac $F^{s}_{1} (t)$ and Pauli $F^{s}_{2} (t)$
strange nucleon FFs
\begin{equation}
   \langle p^{'}|\bar{s}\gamma_{\mu} s|p\rangle=\bar{u}(p^{'})
\left[\gamma_{\mu} F^{s}_{1}(t)+
i{{\sigma_{\mu\nu}q^\nu}\over{2m_N}} F^{s}_{2}(t)\right]u(p)
\label{FFS}
\end{equation}
or in the strange electric and strange magnetic nucleon FFs
\begin{equation}
G_{E}^s(t)=F_{1}^s(t)+\frac{t}{4m_N^2}F_{2}^s(t),\quad
G_{M}^s(t)=F_{1}^s(t)+F_{2}^s(t),
\end{equation}
which, as a consequence of the isospin zero value of the strange
quark, contribute only to the behavior of the isoscalar nucleon
FFs and never to isovector ones.

Further we will predict the nucleon strange FFs behavior by
8-resonance U\&A model and look for distinctive features.

   So, the nucleon Dirac and Pauli strange FFs $F^{s}_{1}(t)$,
$F^{s}_{2}(t)$ can be found just from the behavior of
$F^{I=0}_{1}(t)$, $F^{I=0}_{2}(t)$.
   The  main idea of a prediction of strange nucleon FFs
behavior is based:

\begin{itemize}
\item
on  the $\omega-\phi$ mixing to be valid also for coupling
constants between EM (quark) current and vector meson
\begin{eqnarray}
\frac{1}{f_\omega}=\frac{1}{f_{\omega_0}}\cos
\epsilon-\frac{1}{f_{\phi_0}}\sin\epsilon\\\nonumber
\frac{1}{f_\phi}=\frac{1}{f_{\omega_0}}\sin\epsilon+\frac{1}{f_{\phi_0}}\cos\epsilon
\end{eqnarray}
\item
on the assumption that the quark current of some
flavor couples with universal strength $\kappa$ exclusively
to the vector-meson wave function component of the same flavor
\begin{eqnarray}
<0\mid\bar q_r\gamma_\mu q_r\mid(\bar q_tq_t)_V>=\kappa
m^2_V\delta_{rt}\varepsilon_\mu
\end{eqnarray}
\end{itemize}
which result in the relations
\begin{eqnarray}
\nonumber ({f^{(i)}_{\omega
NN}}/{f^{s}_{\omega}})&=&-\sqrt{6}\frac{\sin{\varepsilon}}
{\sin(\varepsilon+\theta_{0})}({f^{(i)}_{\omega NN}}/{f^{e}_{\omega}})\\
({f^{(i)}_{\phi NN}}/{f^{s}_{\phi}})
&=&-\sqrt{6}\frac{\cos{\varepsilon}}
{\cos(\varepsilon+\theta_{0})}({f^{(i)}_{\phi NN}}/{f^{e}_{\phi}})\\
\nonumber (i&=&1,2)
\end{eqnarray}
where $f^{s}_{\omega},f^{s}_{\phi}$ are strange-current
$\leftrightarrow V=\omega,\phi$ coupling constants and
$\varepsilon=3.7^{0}$ is a deviation from the ideally mixing angle
$\theta_{0}=35.3^{0}$.

So, if one knows from the fit of nucleon FF data free
parameters $(f^{(i)}_{\omega NN}/f^{e}_{\omega})$,
$(f^{(i)}_{\phi NN}/f^{e}_{\phi})$ (i=1,2) of the suitable
model of $F^{I=0}_{1}(t)$, $F^{I=0}_{2}(t)$
\begin{equation}
  F^{I=0}_{i}(t)=f\left[t;({f^{(i)}_{\omega NN}}/{f^{e}_{\omega}}),
({f^{(i)}_{\phi NN}}/{f^{e}_{\phi}})\right] (i=1,2)
\end{equation}
where $f^{(i)}_{\omega NN}, f^{(i)}_{\phi NN}$ are coupling
constants of $\omega$ and $\phi$ to nucleons
and $f^{e}_{\omega}, f^{e}_{\phi}$ are virtual
photon$\leftrightarrow$V=$\omega,\phi$ coupling constants
given by leptonic decay widths $\Gamma(V\rightarrow
e^{+}e^{-})$,
then the unknown free parameters $({f^{(i)}_{\omega
NN}}/{f^{s}_{\omega}}), ({f^{(i)}_{\phi NN}}/{f^{s}_{\phi}})$ of a
strange nucleon FF's model
\begin{equation}
  F^{s}_{i}(t)=\bar{f}\left[t;({f^{(i)}_{\omega NN}}/{f^{s}_{\omega}}),
({f^{(i)}_{\phi NN}}/{f^{s}_{\phi}})\right] (i=1,2) \label{Ms}
\end{equation}
of the same analytic structure are calculated by the relations (22).

Now, why we use 8-resonance U\&A model of nucleon EM structure ?

That follows directly from the derived relations of coupling
constant (strange and EM ratios) where always we are determining
couples of $\omega-\phi$ strange coupling constants
simultaneously. Only 8- 12- etc. resonance U\&A models of nucleon
EM structure fulfill such restrictions, but in no case
10-resonance one.

All known FF properties are contained consistently in the
following specific U\&A models
\begin{eqnarray}
\nonumber && F^{I=0}_{1}[V(t)]
 =(\frac{1-V^{2}}{1-V^{2}_{N}})^{4}\{\frac{1}{2}
 L(V_{\omega''})L(V_{\omega'})+\\
\nonumber && [L(V_{\omega''})L(V_{\omega})
 \frac{(C_{\omega''}-C_{\omega})}
 {(C_{\omega''}-C_{\omega'})}-
 L(V_{\omega'})L(V_{\omega})
 \frac{(C_{\omega'}-C_{\omega})}
 {(C_{\omega''}-C_{\omega'})}-\\
\nonumber && L(V_{\omega''})L(V_{\omega'})]
 (f^{(1)}_{\omega NN}/f^{e}_{\omega})+\\
\nonumber && [L(V_{\omega''})L(V_{\phi})
 \frac{(C_{\omega''}-C_{\phi})}
 {(C_{\omega''}-C_{\omega'})}-
 L(V_{\omega'})L(V_{\phi})
 \frac{(C_{\omega'}-C_{\phi})}
 {(C_{\omega''}-C_{\omega'})}-\\
&& L(V_{\omega''})L(V_{\omega'})]
 (f^{(1)}_{\phi NN}/f^{e}_{\phi})\}
\end{eqnarray}

\begin{eqnarray}
\nonumber &&
F^{I=0}_{2}[V(t)]=(\frac{1-V^2}{1-V^2_N})^6\{L(V_{\omega''})
 L(V_{\omega'})L(V_{\omega})\\
\nonumber && [1-{\frac{C_{\omega}}{(C_{\omega''}-C_{\omega'})}}
 (\frac{(C_{\omega''}-C_{\omega})}{C_{\omega'}}-
 \frac{(C_{\omega'}-C_{\omega})}{C_{\omega''}})]\\
\nonumber && (f^{(2)}_{\omega NN}/f^{e}_{\omega})+
 L(V_{\omega''})L(V_{\omega'})L(V_{\phi})\\
\nonumber && [1-{\frac{C_{\phi}}{(C_{\omega''}-C_{\omega'})}}
 (\frac{(C_{\omega''}-C_{\phi})}{C_{\omega'}}-
 \frac{(C_{\omega'}-C_{\phi})}{C_{\omega''}})]\\
&& (f^{(2)}_{\phi NN}/f^{e}_{\phi})\}
\end{eqnarray}
and

\begin{eqnarray}
\nonumber
&& F^{s}_{1}[V(t)]=(\frac{1-V^{2}}{1-V^{2}_{N}})^{4}\\
\nonumber && \{[L(V_{\omega''})L(V_{\omega})
 \frac{(C_{\omega''}-C_{\omega})}
 {(C_{\omega''}-C_{\omega'})}-
 L(V_{\omega'})L(V_{\omega})
 \frac{(C_{\omega'}-C_{\omega})}
 {(C_{\omega''}-C_{\omega'})}-\\
\nonumber && L(V_{\omega''})L(V_{\omega'})]
 (f^{(1)}_{\omega NN}/f^{s}_{\omega})+\\
\nonumber && [L(V_{\omega''})L(V_{\phi})
 \frac{(C_{\omega''}-C_{\phi})}
 {(C_{\omega''}-C_{\omega'})}-
 L(V_{\omega'})L(V_{\phi})
 \frac{(C_{\omega'}-C_{\phi})}
 {(C_{\omega''}-C_{\omega'})}-\\
&& L(V_{\omega''})L(V_{\omega'})]
 (f^{(1)}_{\phi NN}/f^{s}_{\phi})\}
\end{eqnarray}

\begin{eqnarray}
\nonumber &&
F^{s}_{2}[V(t)]=(\frac{1-V^2}{1-V^2_N})^6\{L(V_{\omega''})
 L(V_{\omega'})L(V_{\omega})\\
\nonumber && [1-{\frac{C_{\omega}}{(C_{\omega''}-C_{\omega'})}}
 (\frac{(C_{\omega''}-C_{\omega})}{C_{\omega'}}-
 \frac{(C_{\omega'}-C_{\omega})}{C_{\omega''}})]\\
\nonumber && (f^{(2)}_{\omega NN}/f^{s}_{\omega})+
 L(V_{\omega''})L(V_{\omega'})L(V_{\phi})\\
\nonumber && [1-{\frac{C_{\phi}}{(C_{\omega''}-C_{\omega'})}}
 (\frac{(C_{\omega''}-C_{\phi})}{C_{\omega'}}-
 \frac{(C_{\omega'}-C_{\phi})}{C_{\omega''}})]\\
&& (f^{(2)}_{\phi NN}/f^{s}_{\phi})\}
\end{eqnarray}
defined each on a four-sheeted Riemann surface with complex
conjugate pairs of resonance poles placed only on the unphysical
sheets, where

\begin{eqnarray*}
\nonumber &&
L(V_r)=\frac{(V_N-V_r)(V_N-V^{\ast}_r)(V_N-1/V_r)(V_N-1/V^{\ast}_r)}
 {(V-V_r)(V-V^{\ast}_r)(V-1/V_r)(V-1/V^{\ast}_r)},\\
&& (r=\omega,\phi,\omega',\omega'')\\
\nonumber &&
C_r=\frac{(V_N-V_r)(V_N-V^{\ast}_r)(V_N-1/V_r)(V_N-1/V^{\ast}_r)}
 {-(V_r-1/V_r)(V^{\ast}_r-1/V^{\ast}_r)},\\
&& (r=\omega,\phi,\omega',\omega'')
\end{eqnarray*}

\begin{equation}
 V(t)=i\frac
 {\sqrt{[\frac{t_{N\bar N}-t^{I=0}_0}{t^{I=0}_0}]^{1/2}+
 [\frac{t-t^{I=0}_0}{t^{I=0}_0}]^{1/2}}-
 \sqrt{[\frac{t_{N\bar N}-t^{I=0}_0}{t^{I=0}_0}]^{1/2}-
 [\frac{t-t^{I=0}_0}{t^{I=0}_0}]^{1/2}}}
 {\sqrt{[\frac{t_{N\bar N}-t^{I=0}_0}{t^{I=0}_0}]^{1/2}+
 [\frac{t-t^{I=0}_0}{t^{I=0}_0}]^{1/2}}+
 \sqrt{[\frac{t_{N\bar N}-t^{I=0}_0}{t^{I=0}_0}]^{1/2}-
 [\frac{t-t^{I=0}_0}{t^{I=0}_0}]^{1/2}}}
\label{ITR}
\end{equation}

\begin{equation}
 V_N=V(t)_{|t=0}; V_r=V(t)_{|t=(m_r-i\Gamma_r/2)^2};
 (r=\omega,\phi,\omega',\omega''),
\label{DEF}
\end{equation}
and $t_{N\bar N}=4m^2_N$ is a square-root branch point
corresponding to $N\bar N$ threshold.

 The expressions for $F^{I=0}_1(t), F^{I=0}_2(t)$ together with
similar expressions for $F^{I=1}_1(t), F^{I=1}_2(t)$ are used:

\begin{itemize}
\item
first to describe Rosenbluth $G_{Ep}$ data in $t<0$ region
together with all other existing nucleon EM FF data with the
result $\chi^2/(ndf)=1.76$
\item
then JLab proton polarization data on $\mu_p
G_{Ep}(t)/G_{Mp}(t)$ in $t<0$ region together with all other
existing nucleon EM FF data with the result $\chi^2/(ndf)=1.34$.
\end{itemize}

The results for $G_{Ep}(t)$ $t<0$ are presented in Fig.8

\begin{figure}[htb]
\centerline{\includegraphics[width=0.55\textwidth]{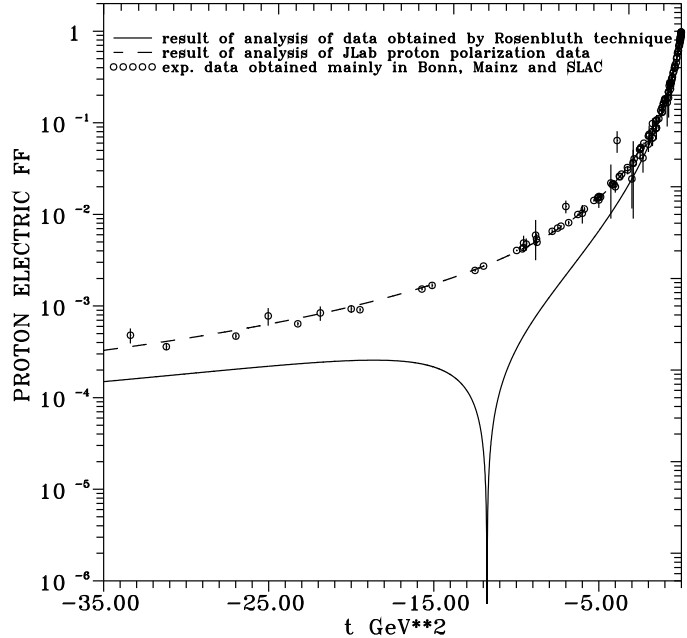}}
\caption{Results of the analysis of SLAC and JLab data by
8-resonance U\&A model}
\end{figure}

   They are similar to the results of the analysis with 10-resonance
model, only the zero is shifted from $t=-13 GeV^2$ to $t=-12 GeV^2$.

\begin{figure}[htb]
\centerline{\includegraphics[width=0.55\textwidth]{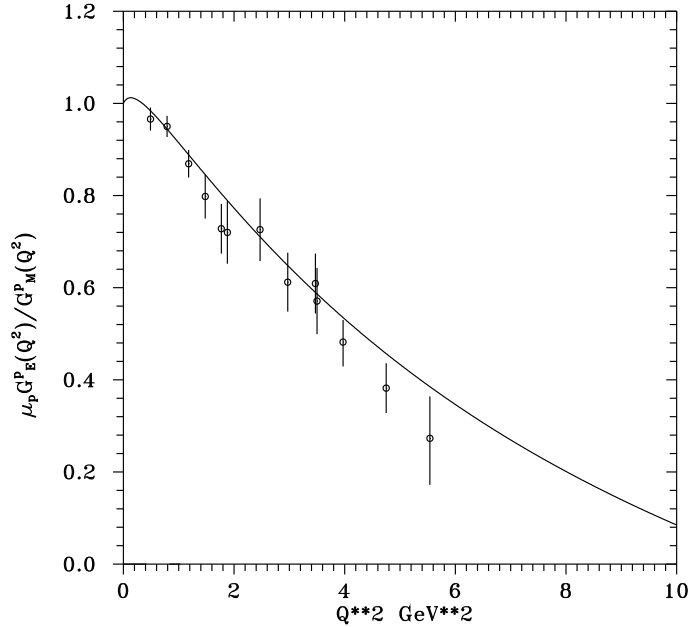}}
\caption{Description of the JLab proton polarization data by
8-resonance U\&A model}
\end{figure}

 Also a perfect description of the JLab proton polarization
data is achieved (see Fig.9).

\begin{figure}[htb]
\centerline{\includegraphics[width=0.45\textwidth]{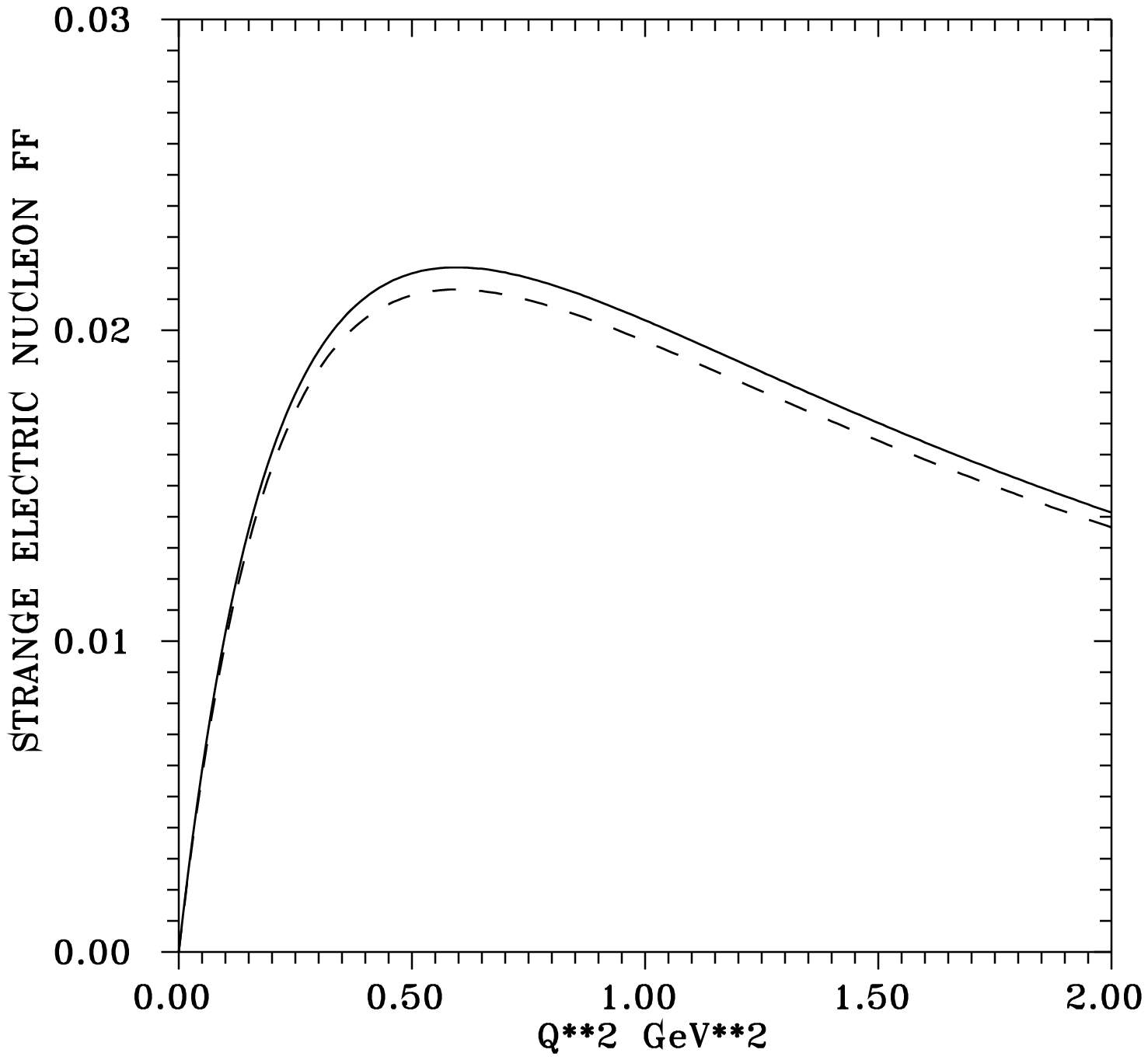}
\qquad
\includegraphics[width=0.45\textwidth]{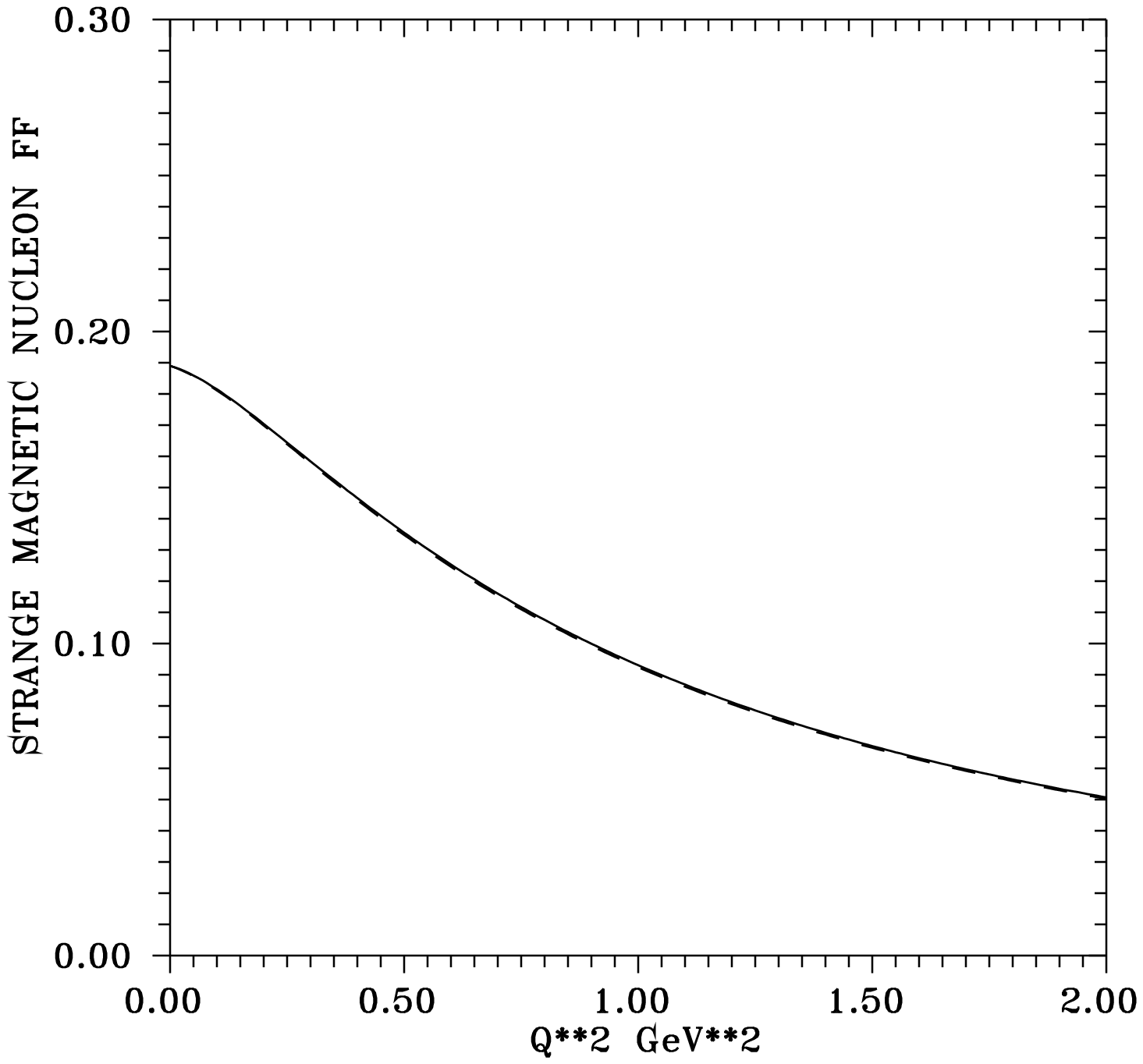}}
\caption{Theoretical prediction of strange electric and magnetic
form factors.}
\end{figure}

Now, calculating $({f^{(i)}_{\omega NN}}/{f^{s}_{\omega}}),
({f^{(i)}_{\phi NN}}/{f^{s}_{\phi}})$ according to the prescribed
procedure and substituting them into the U\&A model of strange
nucleon FFs, one obtains predictions for behaviors of
$G^s_{EN}(Q^2)$ and $G^s_{MN}(Q^2)$ as presented in Fig.10.

As one can see from Fig.10b, a reasonable value of the
strangeness nucleon magnetic moment is predicted $\mu_s=+0.19
[\mu_N]$.

The behavior of strange nucleon FFs doesn't feel too much the
difference in contradicting behaviors of $G_{Ep}(t)$ in space-like
region.

  A reasonable description of the recent data on the combination
  $G^s_E(Q^2)+\eta(Q^2)G^s_M(Q^2)$ for
  $0.12 GeV^2<Q^2<1.0 GeV^2$ is achieved (see Fig.11) \cite{Arm}

\begin{figure}[htb]
\centerline{\includegraphics[width=0.55\textwidth]{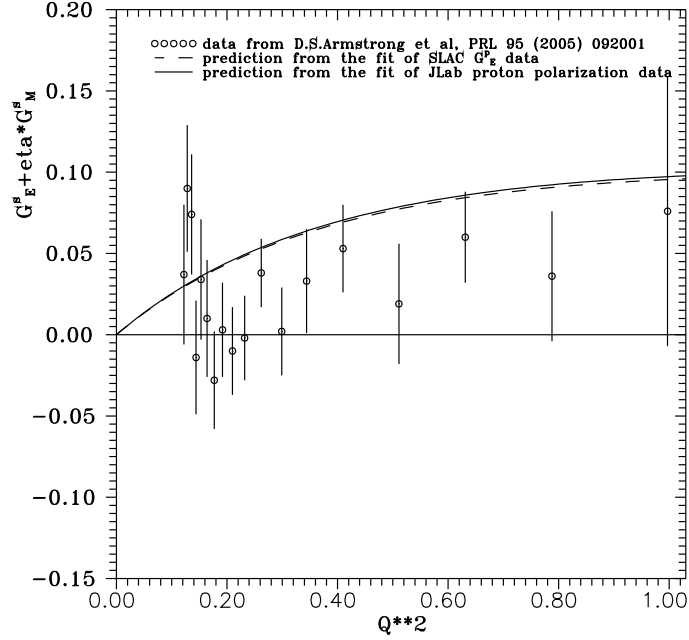}}
\caption{Prediction for the behavior of the combination of strange
nucleon form factors $G^s_E(Q^2)+\eta(Q^2)G^s_M(Q^2)$ by 8-resonance
U\&A model}
\end{figure}

\section{ Deuteron EM structure functions data as a judge between
contradicting $G^p_E(Q^2)$ data}

The cross-section of elastic electron scattering on deuteron is
\begin{eqnarray}
\nonumber \frac{d\sigma^{lab}(e^-D\to
e^-D)}{d\Omega}=\frac{\alpha^2}{4E^2}\frac{\cos^2(\theta/2)}{\sin^4(\theta/2)}
\frac{1}{1+(\frac{2E}{m_D})\sin^2(\theta/2)}.\nonumber
\end{eqnarray}
\begin{equation}
\left[A(t)+B(t)\tan^2(\theta/2)\right]
\end{equation}

Similarly to the nucleons one can draw out from previous formula the
data on $A(Q^2)$ and $B(Q^2)$ as presented in Fig. 12.

\begin{figure}[htb]
\centerline{\includegraphics[width=0.45\textwidth]{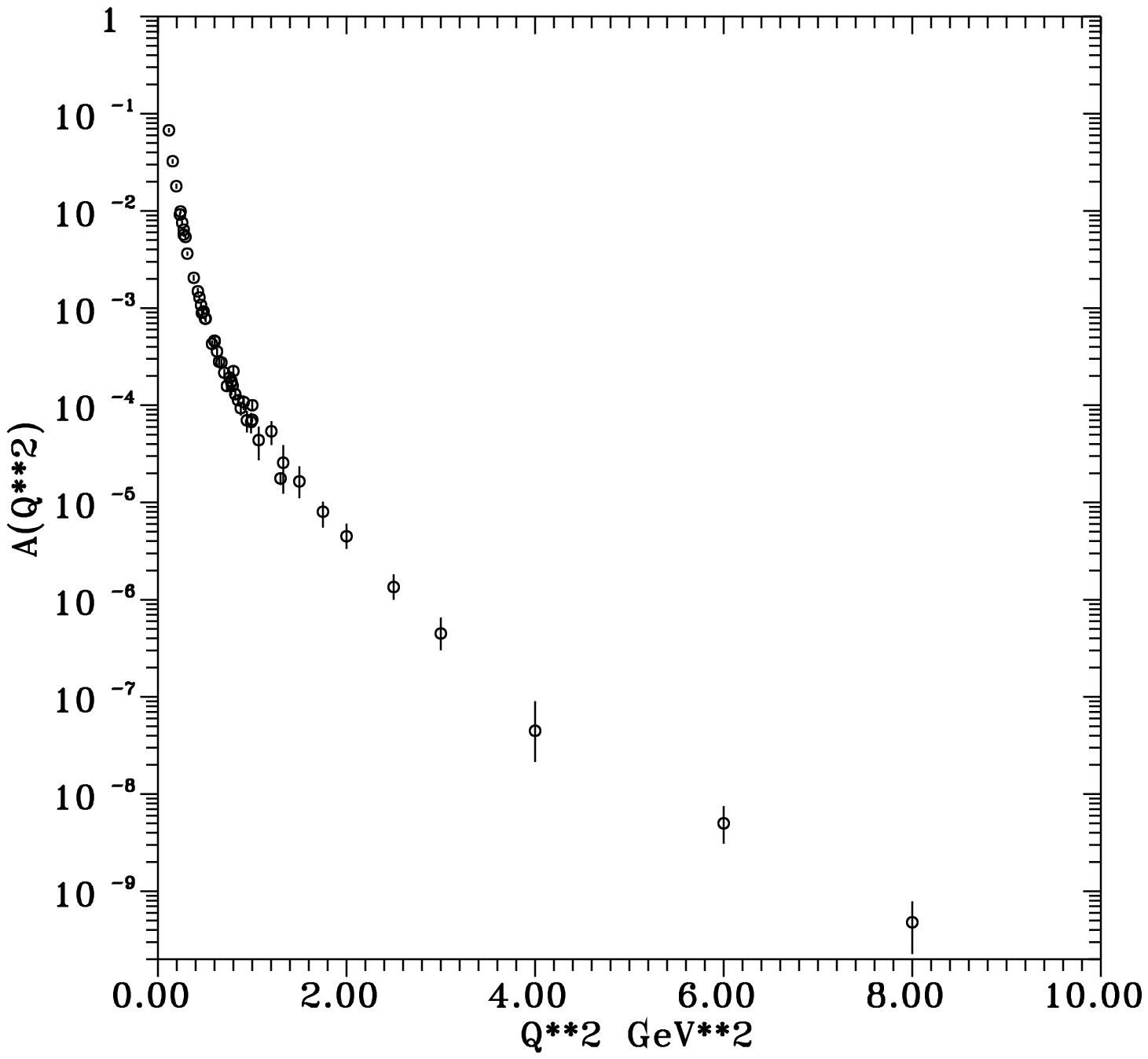}
\qquad
\includegraphics[width=0.45\textwidth]{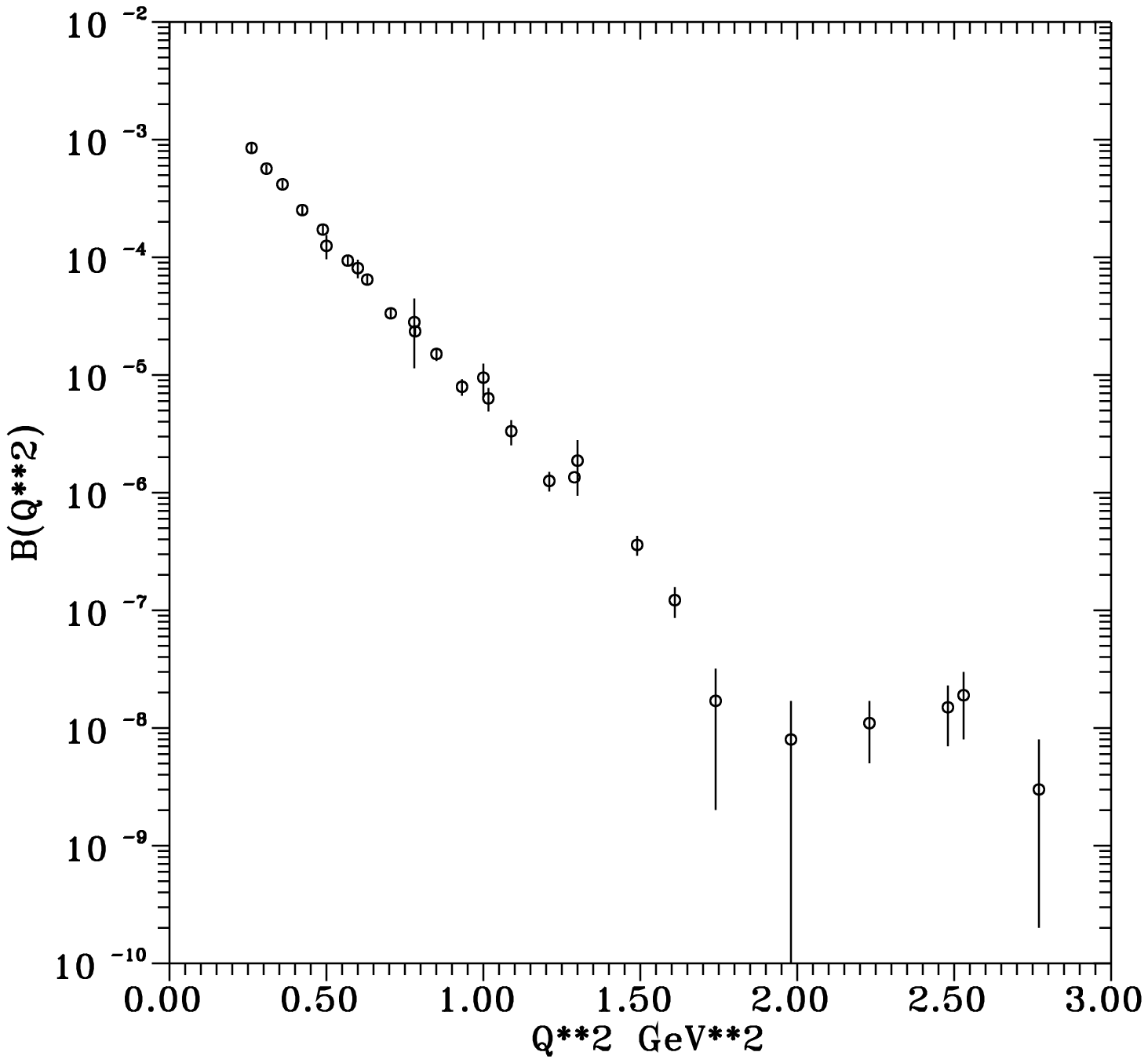}}
\caption{Experimental data on deuteron EM structure functions.}
\end{figure}

   The dependence of $A(Q^2)$ and $B(Q^2)$ on deuteron EM FF's is
given by the relations
\begin{eqnarray*}
A(Q^2)&=&G^2_C(Q^2)+\frac{8}{9}\eta^2 G^2_Q(Q^2)+\frac{2}{3}\eta G^2_M(Q^2)\\
B(Q^2)&=&\frac{4}{3}\eta(1+\eta)G^2_M(Q^2)\\
\eta&=&Q^2/(4m^2_D)
\end{eqnarray*}

In the non-relativistic limit deuteron EM FFs can be
expressed through the iso-scalar parts ($s\sim I=0$) of the
nucleon electric and magnetic FFs
$G^s_E=G^p_E+G^n_E$ and $G^s_M=G^p_M+G^n_M$
\begin{eqnarray*}
G_C&=&G^s_E D_C;\\
G_Q&=&G^s_E D_Q;\\
G_M&=&\frac{m_D}{2m_p}(G^s_MD_M+G^s_E D_E)
\end{eqnarray*}

with

\begin{eqnarray*}
D_C(Q^2)&=&\int_{0}^\infty dr[u^2(r)+w(r)]j_0(qr/2)\\
D_Q(Q^2)&=&\frac{3}{\sqrt{2}\eta} \int_{0}^\infty dr
w(r)[u(r)-w(r)/\sqrt{8}]j_2(qr/2)\\
D_M(Q^2)&=&\int_{0}^\infty dr([2u^2(r)-w^2(r)]j_0(qr/2)+\\
&&+[\sqrt{2}u(r)w(r)+w^2(r)]j_2(qr/2))\\
D_E(Q^2)&=&\frac{3}{2}\int_{0}^\infty dr
w^2(r)[j_0(qr/2)+j_2(qr/2)]
\end{eqnarray*}
and $u(r),w(r)$ are the reduced S- and D-state wave-functions.

   The set of FFs with the new behavior of $G^p_E(Q^2)$ from the
JLab polarization experiments gives a better $\chi^2=1404$
in a description of $A(Q^2)$ and $B(Q^2)$ than the set of FFs
obtained by Rosenbluth technique $\chi^2=2640$.

\section{ Conclusions}

\begin{itemize}
\item
We have presented a manifestation of two contradicting behaviours
of $G^p_E(Q^2)$ in various physical phenomena.
\item
Some of them indicate that the new behavior from the JLab
proton polarization experiment with the zero around $t=-12 GeV^2$
seems to be correct.
\item
However, the source of inconsistency is still left to be
confused.
\end{itemize}

\end{document}